\documentclass[reprint,
superscriptaddress,
amsmath,
amssymb,
aps]{revtex4-2}
\usepackage{multirow}
\usepackage{siunitx} 
\usepackage{braket}
\usepackage{mathtools}
\usepackage{graphicx}
\usepackage{color}
\usepackage{ulem}

\newcommand{\vect}[1]{\mathbf{#1}}

\newcommand{\abs}[1]{\left\lvert #1 \right\rvert}
\newcommand{\figref}[1]{Fig.~\ref{#1}}
\newcommand{\eqnref}[1]{Eq.~(\ref{#1})}

\definecolor{pu}{RGB}{200,50,200}
\definecolor{gr}{RGB}{0,150,0}
\definecolor{bl}{RGB}{68,34,200}
\definecolor{re}{RGB}{200,34,68}
\definecolor{ye}{RGB}{255,165,0}

\begin{document}

\title{A stabilization mechanism for many-body localization in two dimensions}

\author{D. C. W. Foo}
 \email{c2ddfcw@nus.edu.sg}
\affiliation{Centre for Advanced 2D Materials, 
National University of Singapore, 6 Science Drive 2, Singapore 117546}
\author{N. Swain}
\affiliation{Centre for Advanced 2D Materials, 
National University of Singapore, 6 Science Drive 2, Singapore 117546}
\affiliation{MajuLab, International Joint Research Unit IRL 3654, CNRS, Universit\'e C\^ote d'Azur, Sorbonne Universit\'e, National University of Singapore, Nanyang Technological University, Singapore}
\author{P. Sengupta}
\affiliation{Centre for Advanced 2D Materials, 
National University of Singapore, 6 Science Drive 2, Singapore 117546}
\affiliation{School of Physical and Mathematical Sciences, 
Nanyang Technological University, 21 Nanyang Link, Singapore 637371}
\author{\mbox{G. Lemari\'e}}
\affiliation{MajuLab, International Joint Research Unit IRL 3654, CNRS, Universit\'e C\^ote d'Azur, Sorbonne Universit\'e, National University of Singapore, Nanyang Technological University, Singapore}
\affiliation{Centre for Quantum Technologies, National University of Singapore, Singapore 117543}
\affiliation{Laboratoire de Physique Th\'{e}orique, Universit\'{e} de Toulouse, CNRS, UPS, France}
\author{S. Adam}
\affiliation{Centre for Advanced 2D Materials, 
National University of Singapore, 6 Science Drive 2, Singapore 117546}
\affiliation{Department of Materials Science and Engineering, 
National University of Singapore, 9 Engineering Drive 1, 
Singapore 117575}
\affiliation{Yale-NUS College, 16 College Ave West, Singapore 138527}
\affiliation{Department of Physics, Faculty of Science, 
National University of Singapore, 2 Science Drive 3, Singapore 117542}

\begin{abstract}
Experiments in cold atom systems see almost identical signatures of many body localization (MBL) in both one-dimensional ($d=1$) and two-dimensional ($d=2$) systems despite the thermal avalanche hypothesis showing that the MBL phase is unstable for $d>1$. Underpinning the thermal avalanche argument is the assumption of exponential localization of local integrals of motion (LIOMs). In this work we demonstrate that addition of a confining potential -- as is typical in experimental setups -- allows a non-interacting disordered system to have super-exponentially (Gaussian) localized wavefunctions, and an interacting disordered system to undergo a localization transition. Moreover, we show that Gaussian localization of MBL LIOMs shifts the quantum avalanche critical dimension from $d=1$ to $d=2$, potentially bridging the divide between the experimental demonstrations of MBL in these systems and existing theoretical arguments that claim that such demonstrations are impossible.
\end{abstract}

\maketitle

{\it Introduction: }\label{sec:intro}
The study of disordered systems has borne rich discussion and novel phenomena ever since Anderson's seminal work~\cite{Anderson_1958} and the subsequent theoretical observation that {\em all} single particle eigenstates of non-interacting, time-reversal symmetric systems in one and two dimensions are localized in the presence of 
disorder~\cite{anderson_loc_RMP_2008}. Of particular interest in recent years is the phenomenon of Many--Body Localization (MBL), wherein strong disorder drives localization of the entire eigenspectrum in the presence of interactions. MBL has since been subject to intense investigation due to both fundamental and practical reasons~\cite{Polyakov2005,Basko2006,Nandkishore2015,Laflorencie2018,Serbyn2019}. While its existence in 1D is accepted because of good agreement among numerical~\cite{Luitz2015,Serbyn2016,Khemani2016,Lim2016}, analytical~\cite{Imbrie2017} and experimental~\cite{schreiber2015,Smith2016} work, the situation in 2D remains contentious. On one hand, experimental~\cite{mblopt2,mblopt1,mblopt3} and numerical~\cite{2dmblnum1,2dmblnum2,2dmblnum3,2dmblnum4,2dmblnum5,2dmblnumqmc} signatures of MBL in 2D are almost identical to those in 1D~\cite{schreiber2015}, but on the other the thermal avalanche hypothesis (TAH)~\cite{avaltheory,avalnum} posits that MBL cannot exist in any system of dimension greater than 1. However, the TAH relies strongly on the exponential localization of local integrals of motion (LIOMs)~\cite{abanin2013,Abanin2015,liomreview}, an assumption that, as we show below, may be broken on more careful treatment of the disordered potential in these many-body systems.

\begin{figure}[t]
	\centering
	\includegraphics[width=0.9\linewidth]{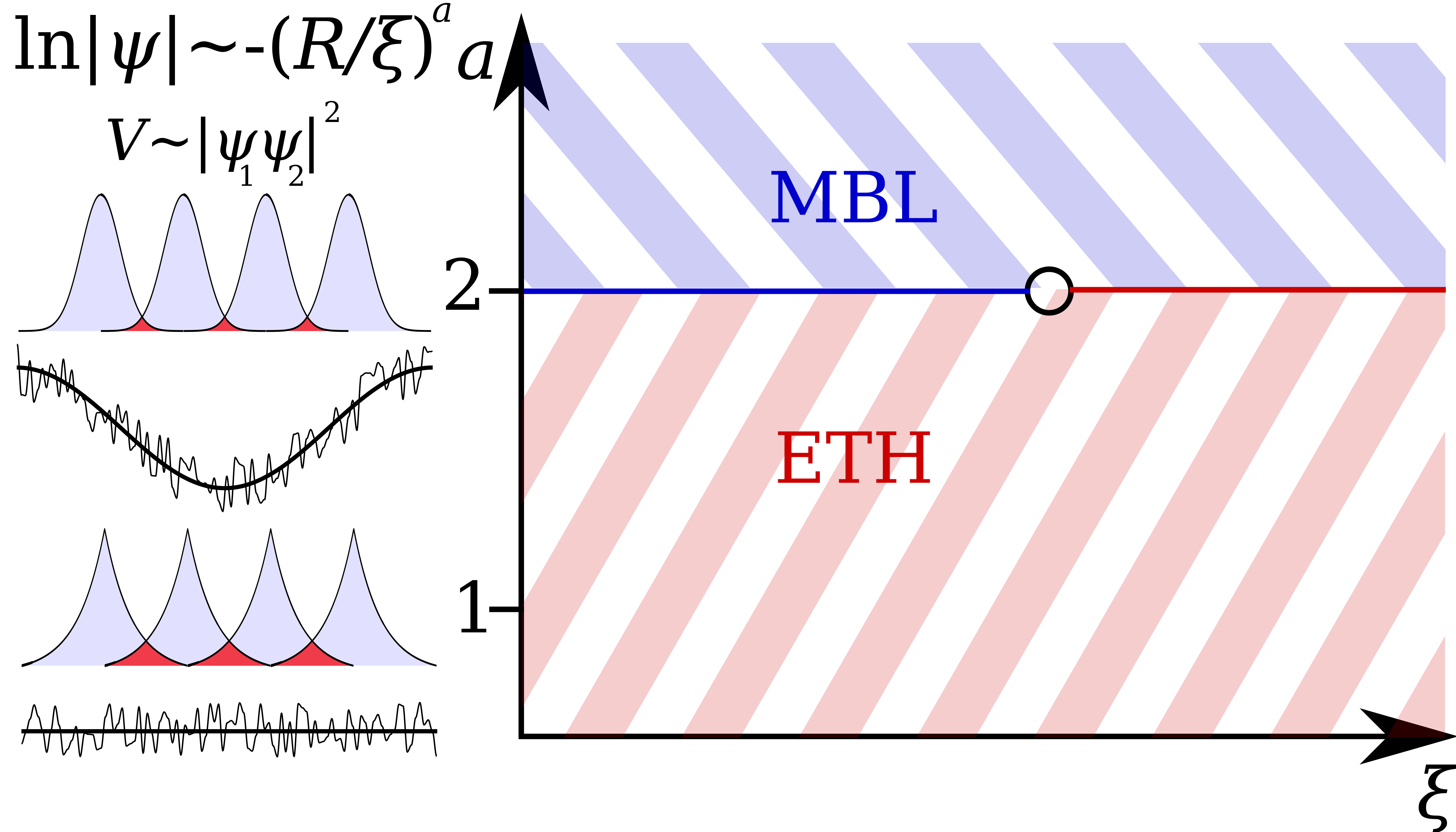}
	\caption{The thermal avalanche hypothesis phase diagram in two dimensions for the many body localization to eigenstate thermalization hypothesis transition as a function of localization length, $\xi$ and the local integral of motion (LIOM) shape parameter $a$, where $a=2$ is critical. LIOM shapes (shaded blue) at both $a=1$ and $a=2$ are shown on the left, with sample localization potentials (black lines) producing those LIOMs shown below each. The confining potential is shown along with the total potential as a guide to the eye. The overlap of adjacent states (shaded red) highlight the qualitative difference between Gaussian and exponential localization of the LIOMs.}
	\label{setup}
\end{figure}

In this Letter, we show that a confining potential, always present in experiments, affects the MBL transition by stabilizing the localized phase. We argue that this is a consequence of super-exponential localization mediated by the confining potential. 

We begin with a brief overview of the TAH, noting in particular the main assumption, which if broken then allows MBL to occur in 2D. We then present evidence of super-exponential (Gaussian) localization in the non-interacting picture. Using exact diagonalization (ED) of interacting spinless fermion Hamiltonians in a disordered cosine trap, we show how such a trap promotes localization, in defiance of the TAH. Our work challenges the commonly accepted view that thermal avalanches always destroys MBL in 2D.


{\it Thermal avalanche hypothesis: }In a sufficiently large system with uncorrelated disorder, it is inevitable for a region of locally weak disorder to emerge. These rare regions of weak disorder may host ``thermal bubbles", regions where the system is well described by the eigenstate thermalization hypothesis (ETH)~\cite{eth1,eth2}. At the interface between localized and thermal regions, interactions between the thermal bubble and individual LIOMs may trigger the thermalization of those LIOMs, thereby incorporating them into the bubble. It has been noted that the situation is highly asymmetric and the reverse, viz., LIOM's localizing the thermal bubble, rarely happens~\cite{avalasym}. The energy scales governing this thermalization are the bubble--LIOM matrix element $\Gamma= V\sqrt{\delta\rho}$ and the bubble level spacing $\delta$, where $V$ is the interaction strength and $\rho$ is the bubble spectral function, with the thermalization of the spin proceeding if $\Gamma\gg\delta$~\cite{avaltheory,mblktscale}. The interaction decays in accordance with the decay (localization) of LIOMs through the relation $V_{ij}\sim\abs{\psi_i\psi_j}^2\sim\exp(-2\abs{i-j}/\xi)$, where $\psi_i$ is the LIOM at site $i$ and $\xi$ is the localization length.

For every additional LIOM incorporated into the thermal bubble, the level spacing roughly halves, and so $\delta$ decays exponentially with the number of thermalized spins. The number of thermalized spins itself grows algebraically with the bubble size $R$, $\delta(R)=\delta_0^{-2A R^d}$, where $\delta_0$ is the bare bubble spacing, $A$ is a positive geometric constant and $d$ is the system dimension. In this way the thermal avalanche is driven by the ever decreasing bubble level spacing while it is limited by the bubble--LIOM interaction strength. Assuming a form $V(R)=V_0\exp(-(R/\xi)^a)$ for the interaction, where $\xi$ is the localization length and $a$ is a shape parameter, and that the bubble spectral function does not change dramatically with $R$, the general criterion for avalanche propagation at a distance $R$ from a thermal bubble in 2D is $\exp\left(AR^2-\left(R/\xi\right)^a\right)\gg1$~\cite{avaltheory,avalreview,avalnum}. We omit a dimensionless prefactor involving comparison of energy scales. The original formulation~\cite{avaltheory} set $a=1$ as the authors assumed exponentially localized LIOMs, and therefore an exponentially decaying coupling between the thermal bubble and LIOMs.


The study of the TAH and the MBL--ETH transition has grown beyond the simple argument described above, with a wealth of numerical and analytical studies discussing, for example, Kosterlitz--Thouless scaling near the critical point~\cite{mblktscale}, localization of the critical point itself~\cite{avalcrit}, coexistence of localized and thermal regions~\cite{avalcoex}, and the dynamical and transport properties~\cite{avaltrans,avalslow}. However, these build upon the basic description above with the same assumption of $a=1$. 

The phase diagram derived from the avalanche condition in 2D is shown in \figref{setup}. Due to the $R$ dependence of the avalanche criterion and the implicit assumption of exponential localization, $a=1$, it has been argued that thermal avalanches unequivocally destroy MBL in 2D, as the quantity on the left hand side increases without bound beyond a critical $R^*$ for $a<2$. However, exponential localization, $a=1$, relies on assumptions that may be broken in real experiments. For example, we argue here that the presence of a trap potentials likely alters the localization profiles to be Gaussian, $a=2$. Such a change of shape qualitatively changes the behaviour of the criterion in 2D, giving rise to a critical $\xi^*$ below which thermal avalanches cannot propagate and therefore cannot destroy MBL. In such a scenario MBL survives in d=2. 

\begin{figure}[t]
	\centering
	\includegraphics[width=0.9\linewidth]{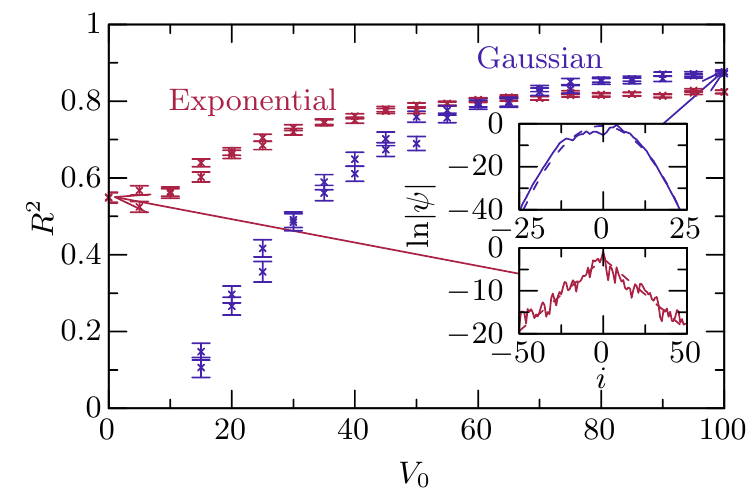}
	\caption{Plot of the coefficient of determination $R^2$ when fitting the absolute value of the lowest energy free state $\abs{\psi}$ to an exponential (\textcolor{re}{red}) or a Gaussian (\textcolor{bl}{blue}) as the depth of a confining potential $V_0$ is increased. Two representative sets of data are shown in each case to show the fit done in two dimensions. A clear evolution from exponential to Gaussian localization on increasing $V_0$ is observed. Insets: Sample $\ln\abs{\psi}$ (solid line) and fitting function (dashed line) for $V_0=0$ (left, \textcolor{re}{red}) and $V_0=100$ (right, \textcolor{bl}{blue})}
	\label{formfit}
\end{figure}

{\it Super-exponential localization: }The concept of exponentially localized single-particle wavefunctions or LIOMs is rooted in two related arguments; first in the Furstenberg theorem~\cite{furstenberg}, which is most applicable in 1D, and second in the forward scattering approximation to the locator expansion~\cite{Anderson_1958,locator1,locator2,forscaapp,liomreview} which can be seen as a mean--field approximation that is more accurate in higher dimensions. Both arguments consider the joint distribution of a product of individual, identically distributed (IID) elements, coming to the conclusion that a Lyapunov exponent naturally emerges characterising the exponential localization of a state. 

The natural question therefore is what happens when these elements are not IID but are instead correlated, as is the case in experiments where the single-particle on-site energy consists of both disorder and confining potential terms~\cite{mblopt2,mblopt1,mblopt3}. This observation opens the possibility of new types of localization and thus the question of what effect the confining potential may have on the localization is of vital importance.

To investigate the possibility of super-exponential localization numerically, we solve the 2D Anderson tight binding model with a confining potential, 
\begin{align}\label{sph}
    H_0=&-\sum_{\langle \vect{i},\vect{j}\rangle}c^\dagger_\vect{i} c^{}_\vect{j}+\sum_\vect{i}(w_\vect{i}+V_\vect{i})n_\vect{i},
\end{align}
where $c^\dagger_{\vect{i}}$ $(c^{}_{\vect{i}})$ creates (annihilates) a spinless fermion at site $\vect{i}=(i_x,i_y)$, $n_\vect{i}$ is the number operator, $L$ is the number of sites in the linear dimension, $w_{\vect{i}}$ is the disordered on-site potential uniformly drawn from $[-W,W]$ and $V_{\vect{i}}=\frac{V_0}{4}\left(\cos\left(\frac{2\pi i_x}{L}\right)+\cos\left(\frac{2\pi i_y}{L}\right)\right)$ is the confining potential. 
Angle brackets denote nearest neighbors (NN), and we impose periodic boundary conditions on both directions. The hopping term sets the energy scale, and we use $W=5$.

We set $L=100$ giving a Hilbert space of $L^2=10^4$ basis states, and look specifically at the eigenstate closest to the center of the spectrum. For each disorder realization, we fit the wavefunction, $\psi$, to both exponential and Gaussian forms in the $x$ and $y$ directions. We then calculate the coefficient of determination, $R^2$, with values closer to 1 indicating a better fit~\cite{stats1}, and average over 400 disorder realizations. \figref{formfit} shows that at fixed disorder strength, as the confining potential depth $V_0$ increases, the wavefunction is better described as having a Gaussian envelope than an exponential one, suggesting that confining potentials may aid in localization by encouraging super-exponential (Gaussian) decay of the wavefunctions. 

\begin{figure}[t]
	\centering
	\includegraphics[width=0.9\linewidth]{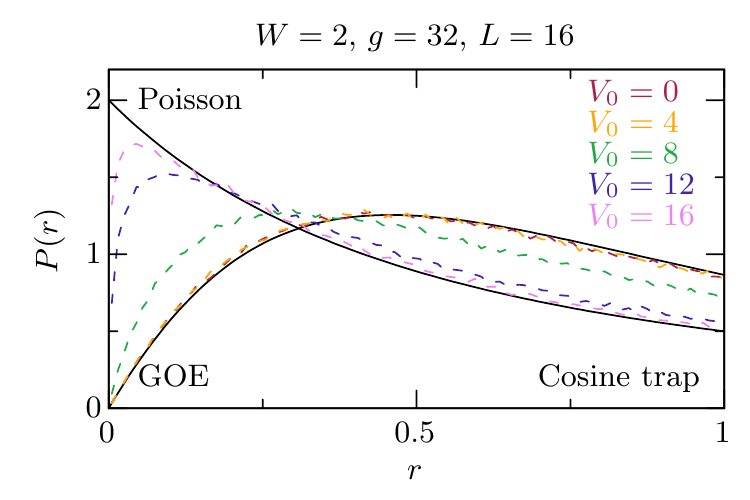}
	\caption{Plot of the distribution, $P(r)$, of level spacing ratios, $r$, for an atomic gas in a cosine trap as trap depth is increased. Without the trap, the system thermalizes despite disorder under the influence of strong, long-range interactions, resulting in a $P(r)$ consistent with the GOE. As trap depth is increased, the system undergoes a localization transition leading to level repulsion and a Poisonnian $P(r)$.}
	\label{lvspace}
\end{figure}

{\it Trap-mediated MBL: }Having demonstrated the change of wavefunction envelope in an Anderson localization context, we now turn on interactions to see how an ordered potential may affect the MBL transition. Owing to the exponential increase in the size of the Hilbert space, we first look at 1D. The Hamiltonian to be discussed is

\begin{align}\label{mblh}
    H=H_0+\frac{g}{L}\sum_{i\neq j}\left(1-\frac{2\abs{i-j}}{L}\right)n_in_j,
\end{align}
where $H_0$ is given by the 1D analogue of \eqnref{sph} with $w_i$ uniformly drawn from $[-W,W]$, $V_i=\frac{V_0}{2}\cos\left(\frac{2\pi i}{L}\right)$, and the second term on the right hand side of \eqnref{mblh} describes the interactions, with $g$ the interaction strength. Infinite range interactions are thought to suppress localization~\cite{longrangembl1,longrangembl2} and so we use these to demonstrate the ability of confining potentials to promote localization. As mentioned, a 1D Hamiltonian is studied to keep the Hilbert space amenable to ED, however, the model is readily generalized to higher dimensions, and we present quantum Monte Carlo (QMC) and ED results for 2D systems in the next section and Supplementary Material respectively. \eqnref{mblh} is analogous to the XXZ spin chain with additional longer range diagonal interactions. For the XXZ spin chain with NN interactions, MBL is thought to occur when $V_0=0$, $g\neq0$ at a critical $W_\mathrm{c}\approx7$~\cite{Luitz2015,xxztransition}. We therefore set $g=32$, $W=2$ to start in the delocalized phase as we investigate the effect of varying $V_0$. 

We probe the localization transition by using ED to obtain the middle 1\% of eigenenergies, repeated over 6400 disorder configurations, and use these to determine the probability distribution $P(r)$ of the ratio of successive energy gaps, $r_i=\min\left(\tilde{r}_i,\frac{1}{\tilde{r}_i}\right)$ with $\tilde{r}_i=(E_{i+2}-E_{i+1})/(E_{i+1}-E_i)$. For time-reversal symmetric Hamiltonians such as \eqnref{mblh}, $P(r)$ is expected to follow the Gaussian orthogonal ensemble (GOE) in the thermal delocalized phase and a Poisson distribution in the MBL phase~\cite{lstats1,lstats2,rstats,mblstats}. For $V_0=0$ and $V_0=4$, $P(r)$ closely adheres to that of the GOE before breaking away and transitioning to the Poisson distribution at higher values of $V_0$, seen in the evolution of dashed lines in~\figref{lvspace}. These results clearly indicate a transition from thermal to localized mediated by the trap depth.

These results bear some resemblance to previous work on Stark--MBL~\cite{starkmbl1,starkmbl3,smbldisfreeharm}, though we note that we use long range interactions and work with periodic rather than open boundary conditions, and therefore in the $W=0$ limit the system does not admit localized solutions. This is in contrast to the Wannier--Stark localization that precedes Stark--MBL~\cite{wannierloc,starkmbl2}. The localization observed here therefore requires both disorder and appropriate confining potential.

\begin{figure}[t]
	\centering
	\includegraphics[width=0.9\linewidth]{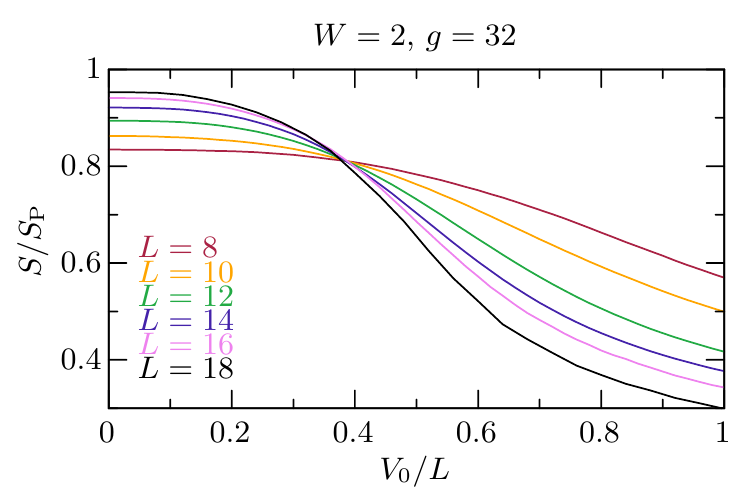}
	\caption{Entanglement entropy of half the system against trap aspect ratio with increasing system size. Shallow traps permit a thermal phase while deep traps promote localization, with a crossover aspect ratio of $(V_0/L)_\mathrm{c}\approx0.4$, corresponding to a vanishing trap frequency $\omega=\pi\sqrt{V_0}/L$ in the thermodynamic limit. The errors of the curves are within 1\% and are omitted for clarity.}
	\label{ee}
\end{figure}

{\it Finite-size effects: }An alternate probe to identify the MBL transition is the entanglement entropy, defined as $S=-\mathrm{Tr}_A\left(\rho_A\log\rho_A\right)$ where $\rho_A=\mathrm{Tr}_B\ket{\psi}\bra{\psi}$ is the partial density matrix of subsystem $A$ when the entire system is in the state $\psi$. Subsystem $B$ is the complement of $A$.
The trapping potential, $V_i$, removes the ``translationally symmetric" freedom to choose subsystems $A$ and $B$; here we specifically choose both to span a half period of $V_i$ from maxima to minima. Results are shown in~\figref{ee}, where we average over the middle 1\% or 100 eigenstates, whichever is fewer, and 6400 disorder configurations. We furthermore normalize by the Page value~\cite{pageentropy}, $S_\mathrm{P}=(L\log2-1)/2$ to get a figure of merit between 0 and 1, where 0 indicates MBL and 1 indicates purely thermal behaviour. The normalized entanglement entropy is plotted against the aspect ratio of the trap potential, $V_0/L$, for various values of $L$ to get a sense of the transition point and the magnitude of finite size effects. The results are in agreement with those for $P(r)$, indicating a thermal to localized transition on increasing $V_0$ with a critical aspect ratio of $(V_0/L)_\mathrm{c}\approx0.4$ and the transition sharpening as $L$ increases. 

Further insight may be obtained by considering the harmonic approximation of the confining potential about its minimum, $V_i\approx \pi^2V_0i^2/L^2$ for $i$ near $L/2$. Identifying this potential with that of the quantum harmonic oscillator gives a trap frequency $\omega=\pi\sqrt{V_0}/L$ which scales as $1/\sqrt{L}$ at the critical point, indicating that in the thermodynamic limit, the critical trap frequency tends to 0. These results suggest that the trap-mediated localization persists in the thermodynamic limit.


{\it 2D numerics: }We use the stochastic series expansion (SSE) QMC technique to investigate the 2D analog of \eqnref{mblh} with NN interactions, 

\begin{align}\label{mblh2dnn}
    H=&H_0+g\sum_{\langle\vect{i},\vect{j}\rangle}n_{\vect{i}}n_{\vect{j}},
\end{align}
with $w_{\vect{i}}$ uniformly distributed in $[-W,W]$ and $V_{\vect{i}}=\frac{V_0}{4}\left(\cos\left(\frac{2\pi i_x}{L}\right)+\cos\left(\frac{2\pi i_y}{L}\right)\right)$.

\begin{figure}[t]
	\centering
	\includegraphics[width=0.9\linewidth]{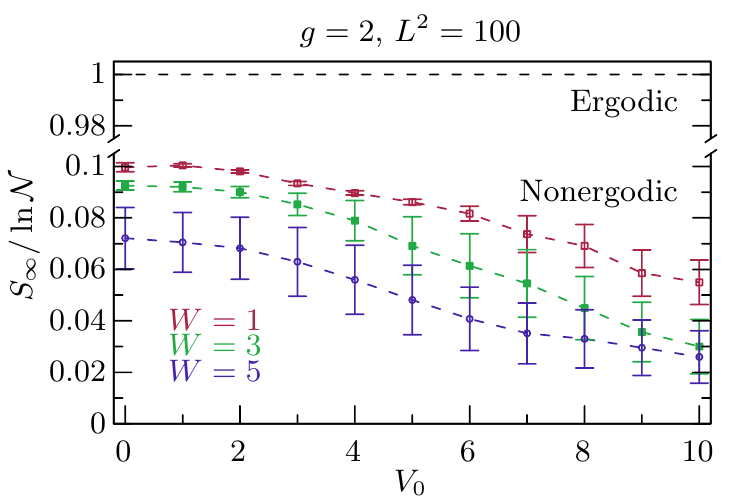}
	\caption{Participation entropy of the ground state of 2D NN-interacting spinless fermions with on-site disorder in a cosine confining potential, calculated by SSE QMC. The participation entropy decreases with increasing trap depth, indicating increased nonergodicity. The black dashed line at $S_\infty/\ln\mathcal{N}=1$ indicates ergodic behaviour.}
	\label{qmc}
\end{figure}

QMC algorithms are restricted to probing only ground state properties of a Hamiltonian, $H$. This hurdle may be mitigated by using the Eigenstate to Hamiltonian construction~\cite{ehc1} to find a mapped Hamiltonian, $\tilde{H}$, with the same interaction strength and trap depth but different disorder configuration, hosting the ground state of the parent Hamiltonian as an excited state of $\tilde{H}$. The ground state properties of $H$, estimated with SSE QMC,therefore describe the excited state properties of $\tilde{H}$ of the same form~\cite{2dmblnumqmc,ehc2,ehc3}. Simulating a 2D system of $10\times10$ lattice sites, we calculate the participation entropy, 
\begin{align}
    S_\infty=\lim_{q\to\infty}\frac{1}{1-q}\ln\left(\sum_i\abs{\braket{\psi|\phi_i}}^{2q}\right),
\end{align}
where $\psi$ is the many-body ground state and $\phi_i$ is the $i$-th Fock basis state. The value $S_\infty/\ln \mathcal{N}$, with $\mathcal{N}$ the Hilbert space dimension, denotes the multifractal dimension of the state in the Fock space, whose finiteness is a characteristic signature of the MBL phase, while it is equal to 1 in the ergodic phase. We see in~\figref{qmc} that the multifractal dimension decreases as $V_0$ increases, indicating increasingly nonergodic behavior induced by the confining potential, thus confirming the 1D results.

{\it Discussion and conclusions: }We have revisited the TAH and noted in particular that the underlying assumption of exponentially localized LIOMs may be broken under application of trap potentials, relevant for real experiments~\cite{mblopt1,mblopt2,mblopt3}, and that the localization decay can be made Gaussian. This observation directly challenges the assertion that MBL is generically unstable to thermal avalanches in dimension greater than 1 and opens up the possibility for stable MBL in 2D, as has previously been reported in experimental~\cite{mblopt2,mblopt1,mblopt3} and numerical~\cite{2dmblnum1,2dmblnum2,2dmblnum3,2dmblnum4,2dmblnum5,2dmblnumqmc} studies.

We have further demonstrated that the addition of such a potential term to an MBL Hamiltonian may trigger an MBL transition in a parameter regime that would otherwise host a thermal phase, and that such a transition persists in the thermodynamic limit. Nevertheless, the reader may be concerned about whether the observed stabilization of MBL in 2D are truly the result of Gaussian localization or might have some other physical origin brought on by the addition of the confining potential. Two alternative possibilities are fragmenting of the Hilbert space through emergence of new conserved quantities, and the rescaling of effective hopping amplitudes as the confining potential impedes particle motion. We show in the Supplementary Material that neither of these explanations account for the observed localization. Our results point to external trap potentials and the concomitant Gaussian localization being the most likely stabilization mechanism for MBL in 2D and in principle able to overcome the TAH. 

Potential avenues of further work include the study of the effect of trap potentials on the dynamics of disordered systems, and experiments to verify stability of the MBL phase against thermal avalanches by changing the trap shape or depth. For example, a 2D system confined with a cosine, or locally quadratic, potential in one direction and a uniform square well trap~\cite{uniformtrap} in the orthogonal direction could give an anisotropic avalanche propagation. 



The authors would like to thank Sreedevi Athira Krishnan and Kai Dieckmann for their helpful discussions and acknowledge the financial support of Singapore Ministry of Education AcRF Tier 2 grants MOE2017-T2-1-130 and MOE2019-T2-2-118, and the Singapore National Research Foundation Investigator Award (NRF-NRFI06-2020-0003). Gabriel Lemari\'{e} acknowledges the support of the projects GLADYS ANR-19- CE30-0013 and MANYLOK ANR-18-CE30-0017 of the French National Research Agency (ANR), by the Singapore Ministry of Education Academic Research Fund
Tier I (WBS No. R-144- 000-437-114).

\bibliographystyle{unsrt}
\bibliography{bibo}

\begin{thebibliography}{10}

\bibitem{Anderson_1958}
{P.~W.~Anderson}.
\newblock {\em Phys.~Rev.}, 109(5):1492--1505, 1958.

\bibitem{anderson_loc_RMP_2008}
{F.~Evers and A.~D.~Mirlin}.
\newblock {\em Rev.~Mod.~Phys.}, 80(4):1355--1417, 2008.

\bibitem{Polyakov2005}
{I.~V.~Gornyi, A.~D.~Mirlin, and D.~G.~Polyakov}.
\newblock {\em Phys.~Rev.~Lett.}, 95(206603), 2005.

\bibitem{Basko2006}
{D.~M.~Basko, I.~L.~Aleiner and B.~L.~Altshuler}.
\newblock {\em Ann.~Phys.~(N.~Y.)}, 321(5):1126--1205, 2006.

\bibitem{Nandkishore2015}
{R.~Nandkishore and D.~A.~Huse}.
\newblock {\em Annu.~Rev.~Condens.~Matter~Phys.}, 6(1):15--38, 2015.

\bibitem{Laflorencie2018}
{F.~Alet and N.~Laflorencie}.
\newblock {\em C.~R.~Phys.}, 19(498), 2018.

\bibitem{Serbyn2019}
{D.~A.~Abanin, E.~Altman, I.~Bloch, and M.~Serbyn}.
\newblock {\em Rev.~Mod.~Phys.}, 91(021001), 2019.

\bibitem{Luitz2015}
{D.~J.~Luitz, N.~Laflorencie and F.~Alet}.
\newblock {\em Phys.~Rev.~B}, 91(8):081103(R), 2015.

\bibitem{Serbyn2016}
{M.~Serbyn and J.~E.~Moore}.
\newblock {\em Phys.~Rev.~B}, 93(4):041424(R), 2016.

\bibitem{Khemani2016}
{V.~Khemani, F.~Pollmann and S.~L.~Sondhi}.
\newblock {\em Phys.~Rev.~Lett.}, 116(24):247204, 2016.

\bibitem{Lim2016}
{S.~P.~Lim and D.~N.~Sheng}.
\newblock {\em Phys.~Rev.~B}, 94(4):045111, 2016.

\bibitem{Imbrie2017}
{J.~Z.~Imbrie, V.~Ros and A.~Scardicchio}.
\newblock {\em Annalen~der~Physik}, 529(7):1600278, 2017.

\bibitem{schreiber2015}
{M.~Schreiber, S.~S.~Hodgman, P.~Bordia, H.~P.~L{\"{u}}schen, M.~H.~Fischer,
  R.~Vosk, E.~Altman, U.~Schneider and I.~Bloch}.
\newblock {\em Science}, 349(6250):842--845, 2015.

\bibitem{Smith2016}
{J.~Smith, A.~Lee, P.~Richerme, B.~Neyenhuis, P.~W.~Hess, P.~Hauke, M.~Heyl,
  D.~A.~Huse and C.~Monroe}.
\newblock {\em Nat.~Phys.}, 12(10):907, 2016.

\bibitem{mblopt2}
{J.~Choi, S.~Hild, J.~Zeiher, P.~Schau{\ss}, A.~Rubio-Abadal, T.~Yefsah,
  V.~Khemani, D.~A.~Huse, I.~Bloch and C.~Gross}.
\newblock {\em Science}, 352(1547), 2016.

\bibitem{mblopt1}
{S.~S.~Kondov, W.~R.~McGehee, W.~Xu, and B.~DeMarco}.
\newblock {\em Phys.~Rev.~Lett.}, 114(083002), 2015.

\bibitem{mblopt3}
{P.~Bordia, H.~L\"{u}schen, S.~Scherg, S.~Gopalakrishnan, M.~Knap,
  U.~Schneider, and I.~Bloch}.
\newblock {\em Phys.~Rev.~X}, 7(041047), 2017.

\bibitem{2dmblnum1}
{T.~B.~Wahl, A.~Pal, and S.~H.~Simon}.
\newblock {\em Nature~Physics}, 15(164), 2019.

\bibitem{2dmblnum2}
{A.~Kshetrimayum, M.~Goihl, and J.~Eisert}.
\newblock {\em Phys.~Rev.~B}, 102(235132), 2020.

\bibitem{2dmblnum3}
{H.~Th\'{e}veniaut, Z.~Lan, G.~Meyer, F.~Alet}.
\newblock {\em Phys.~Rev.~Research}, 2(033154), 2020.

\bibitem{2dmblnum4}
{E.~Chertkov, B.~Villalonga, and B.~K.~Clark}.
\newblock {\em Phys.~Rev.~Lett.}, 126(180602), 2021.

\bibitem{2dmblnum5}
{K.~S.~C.~Decker, D.~M.~Kennes, C.~Karrash}.
\newblock {\em arXiv: 2106.12861}, 2021.

\bibitem{2dmblnumqmc}
{H.~Tang, N.~Swain, D.~C.~W.~Foo, B.~J.~J.~Khor, F.~F.~Assaad, S.~Adam,
  P.~Sengupta}.
\newblock {\em arXiv: 2106.08587}, 2021.

\bibitem{avaltheory}
{W.~De Roeck and F.~Huveneers}.
\newblock {\em Phys.~Rev.~B}, 95(155129), 2017.

\bibitem{avalnum}
{I.-D.~Potirniche, S.~Banerjee, and E.~Altman}.
\newblock {\em Phys.~Rev.~B}, 99(205149), 2019.

\bibitem{abanin2013}
{M.~Serbyn, Z.~Papi\'{c}, and D.~A.~Abanin}.
\newblock {\em Phys.~Rev.~Lett}, 111(127201), 2013.

\bibitem{Abanin2015}
{A.~Chandran, I.~H.~Kim, G.~Vidal, and D.~A.~Abanin}.
\newblock {\em Phys.~Rev.~B}, 91(085425), 2015.

\bibitem{liomreview}
{J.~Z.~Imbrie, V.~Ros, A.~Scardicchio}.
\newblock {\em Annalen Der Physik}, 529(1600278), 2017.

\bibitem{eth1}
{J.~M.~Deutsch}.
\newblock {\em Phys.~Rev.~A}, 43(2046), 1991.

\bibitem{eth2}
{M.~Srednicki}.
\newblock {\em Phys.~Rev.~E}, 50(888), 1994.

\bibitem{avalasym}
{D.~J.~Luitz, F.~Huveneers, and W.~De Roeck}.
\newblock {\em Phys.~Rev.~Lett.}, 119(150602), 2017.

\bibitem{mblktscale}
{P.~T.~Dumitrescu, A.~Goremykina, S.~A.~Parameswaran, M.~Serbyn and
  R.~Vasseur}.
\newblock {\em Phys.~Rev.~B}, 99(094205), 2019.

\bibitem{avalreview}
{W.~De Roeck and J.~Z.~Imbrie}.
\newblock {\em Phil.~Trans.~R.~Soc.~A}, 375(20160422), 2017.

\bibitem{avalcrit}
{T.~Thiery, F.~Huveneers, M.~M\"{u}ller and W.~De Roeck}.
\newblock {\em Phys.~Rev.~Lett.}, 121(140601), 2018.

\bibitem{avalcoex}
{P.~J.~D.~Crowley and A.~Chandran}.
\newblock {\em Phys.~Rev.~Research}, 2(033262), 2020.

\bibitem{avaltrans}
{S.~Gopalakrishnan, S.~A.~Parameswaran}.
\newblock {\em Phys.~Rep}, 862(1), 2020.

\bibitem{avalslow}
{E.~V.~H.~Doggen, I.~V.~Gornyi, A.~D.~Mirlin, and D.~G.~Polyakov}.
\newblock {\em Phys.~Rev.~Lett.}, 125(155701), 2020.

\bibitem{furstenberg}
{H.~Furstenberg}.
\newblock {\em Trans.~Amer.~Math.~Soc.}, 108(377), 1963.

\bibitem{locator1}
{D.~M.~Basko, I.~L.~Aleiner, and B.~L.~Altshuler}.
\newblock {\em Annals of Physics}, 321(1126), 2006.

\bibitem{locator2}
{D.~M.~Basko, I.~L.~Aleiner, and B.~L.~Altshuler}.
\newblock {\em Phys.~Rev.~B}, 76(052203), 2007.

\bibitem{forscaapp}
{F.~Pietracaprina, V.~Ros, A.~Scardicchio}.
\newblock {\em Phys.~Rev.~B}, 93(054201), 2016.

\bibitem{stats1}
R.~G.~D. Steel and J.~H. Torrie.
\newblock {\em Principles and Procedures of Statistics with Special Reference
  to the Biological Sciences.}
\newblock McGraw Hill, 1960.

\bibitem{longrangembl1}
{R.~M.~Nandkishore, S.~L.~Sondhi}.
\newblock {\em Phys.~Rev.~X}, 7(041021), 2017.

\bibitem{longrangembl2}
{S.~Nag, A.~Garg}.
\newblock {\em Phys.~Rev.~B}, 99(224203), 2019.

\bibitem{xxztransition}
{M.~Serbyn, Z.~Papi\'{c}, and D.~A.~Abanin}.
\newblock {\em Phys.~Rev.~X}, 5(041047), 2015.

\bibitem{lstats1}
{V.~Oganesyan and D.~A.~Huse}.
\newblock {\em Phys.~Rev.~B}, 75(155111), 2007.

\bibitem{lstats2}
{A.~Pal and D.~A.~Huse}.
\newblock {\em Phys.~Rev.~B}, 82(174411), 2010.

\bibitem{rstats}
{Y.~Y.~Atas, E.~Bogomolny, O.~Giraud, and G.~Roux}.
\newblock {\em Phys.~Rev.~Lett.}, 110(084101), 2013.

\bibitem{mblstats}
{P.~Sierant and J.~Zakrzewski}.
\newblock {\em Phys.~Rev.~B}, 99(104205), 2019.

\bibitem{starkmbl1}
{M.~Schulz, C.~A.~Hooley, R.~Moessner and F.~Pollmann}.
\newblock {\em Phys.~Rev.~Lett.}, 121(040606), 2019.

\bibitem{starkmbl3}
{S.~R.~Taylor, M.~Schulz, F.~Pollmann and R.~Moessner}.
\newblock {\em Phys.~Rev.~B}, 102(054206), 2020.

\bibitem{smbldisfreeharm}
{T.~Chanda, R.~Yao, and J.~Zakrzewski}.
\newblock {\em Phys.~Rev.~Research}, 2(032039(R)), 2020.

\bibitem{wannierloc}
{G.~H.~Wannier}.
\newblock {\em Phys.~Rev.}, 117(432), 1960.

\bibitem{starkmbl2}
{E.~P.~L.~van Nieuwenburg, Y.~Baum, and G.~Refael}.
\newblock {\em PNAS}, 116(19), 2019.

\bibitem{pageentropy}
{D.~N.~Page}.
\newblock {\em Phys.~Rev.~Lett.}, 71(1291), 1993.

\bibitem{ehc1}
{E.~Chertkov and B.~K.~Clark}.
\newblock {\em Phys.~Rev.~X}, 8(031029), 2018.

\bibitem{ehc2}
{X.~l.~Qi and D.~Ranard}.
\newblock {\em {Quantum}}, 3(159), 2019.

\bibitem{ehc3}
{M.~Dupont and N.~Laflorencie}.
\newblock {\em Phys.~Rev.~B}, 99(020202(R)), 2019.

\bibitem{uniformtrap}
{B.~Mukherjee, Z.~Yan, P.~B.~Patel, Z.~Hadzibabic, T.~Yefsah, J.~Struck and
  M.~W.~Zwierlein}.
\newblock {\em Phys.~Rev.~Lett.}, 118(123401), 2017.

\bibitem{univmoire}
{J.~Attig, J.~Park, M.~M.~Scherer, S.~Trebst, A.~Altland, A.~Rosch}.
\newblock {\em 2D~Materials}, 8(044007), 2021.

\end{thebibliography}

\section*{Supplementary Material}

\begin{figure}[h]
	\centering
	\includegraphics[width=0.9\linewidth]{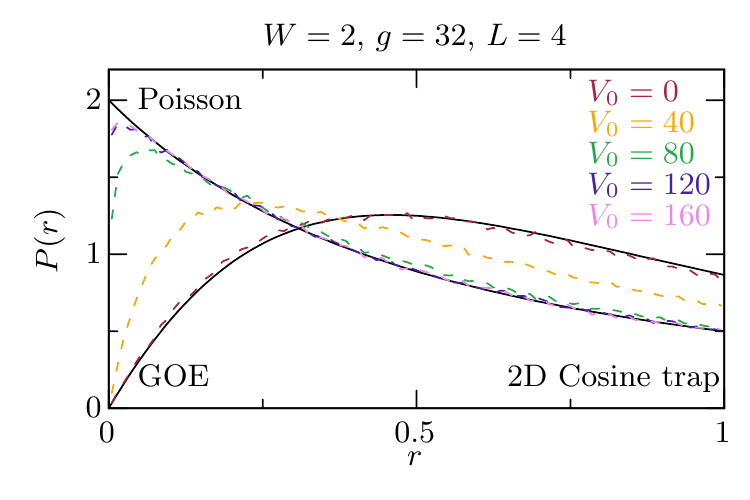}
	\caption{Plot of the distribution, $P(r)$, of level spacing ratios, $r$, for a 2D atomic gas described by \eqnref{mblh2d} as trap depth is increased. Data gathered using ED. Without the trap, the system thermalizes despite disorder under the influence of strong, long-range interactions, resulting in a $P(r)$ consistent with the GOE. As trap depth is increased, the system undergoes a localization transition leading to absence of level repulsion and a Poisonnian $P(r)$.}
	\label{lvspace2d}
\end{figure}

{\it Exact diagonalization results in 2D: }Given the main argument that Gaussian localization may overcome thermalization in 2D, it is prudent to present evidence of MBL in a 2D system, even if the system size accessible to us is small. A straightforward extension of \eqnref{mblh} gives

\begin{align}\label{mblh2d}
    H=&H_0+\frac{g}{L^2}\sum_{\vect{i}\neq\vect{j}}\left(1-\frac{\sqrt{2}\abs{\vect{i}-\vect{j}}}{L}\right)n_\vect{i}n_\vect{j},
\end{align}
with $w_i$ uniformly drawn from $[-W,W]$ and $V_{\vect{i}}=\frac{V_0}{4}\left(\cos\left(\frac{2\pi i_x}{L}\right)+\cos\left(\frac{2\pi i_y}{L}\right)\right)$, which we exactly diagonalize in analogy to \figref{lvspace} for a $4\times4$ system, all other physical and numerical parameters being equal. Results are shown in \figref{lvspace2d}, where the same GOE to Poissonian evolution of $P(r)$ on increasing $V_0$ is seen as in the 1D study of the main text, though the trap depth required here is much higher. This is due to the increased coordination of the hopping and interaction terms promoting thermalization. However, due to the small system size, the lattice model cannot resolve the cosine confining potential and at $L=4$ it looks like a sawtooth wave. There is therefore a need for alternate techniques such as the QMC -- EHC approach that we use in our letter to properly investigate 2D systems.


{\it Multiple wells: }There is no need \emph{a priori} to identify the confining potential wavelength with the system size, and we could for example make the replacement $L\to kL$ for the system size with $k\in\mathbb{Z}^+$ in \eqnref{sph} and maintain the periodic boundary conditions. The quantity $k$ would then count the number of wavelengths spanned by the system. The results obtained for $k>1$ and finite disorder had the wavefunction localized about a single point in space with no obvious periodicity, and were indistinguishable from $k=1$, so we choose $k=1$ for computational expedience. On the contrary, at zero disorder, the eigenstates were periodic in $L$ as expected from Bloch's theorem. We therefore conclude that for periodic systems, the breaking of discrete translational symmetry by inclusion of disorder is necessary for localization about a single point. Such concerns have thus far not been addressed in studies of disorder-free Stark--MBL~\cite{smbldisfreeharm,starkmbl1,starkmbl2,starkmbl3} where open boundary conditions are used. 

{\it Other potentials: }Further investigation into the importance of Gaussian localization envelopes may be conducted by varying the form of the confining potential present in Eqns. (1) and (2). We choose two simple periodic forms, 
\begin{align}
    V_i^{(\mathrm{Saw})}&= V_0\abs{\frac{2i}{L}-1}\label{vsaw}\\V_i^{(\mathrm{Sq})}&=V_0\Theta\left(\frac{2i}{L}-1\right)\label{vsq},
\end{align}
where $\Theta(x)$ is the Heaviside step function and the 2D versions are readily generalized as $V_\vect{i}=\tfrac{1}{2}(V_{i_x}+V_{i_y})$. The sawtooth wave in particular is reminiscent of the constant gradient Stark field seen in Stark--MBL studies~\cite{starkmbl1,starkmbl3,smbldisfreeharm,starkmbl3} but we repeat that the previous studies were performed with open boundary conditions and that in the non-interacting disorder free limit, while the open system admits exponentially localized Wannier--Stark orbitals, the periodic system in contrast does not admit localized solutions as a consequence of Bloch's Theorem. 

\begin{figure}[t]
	\centering
	\includegraphics[width=0.9\linewidth]{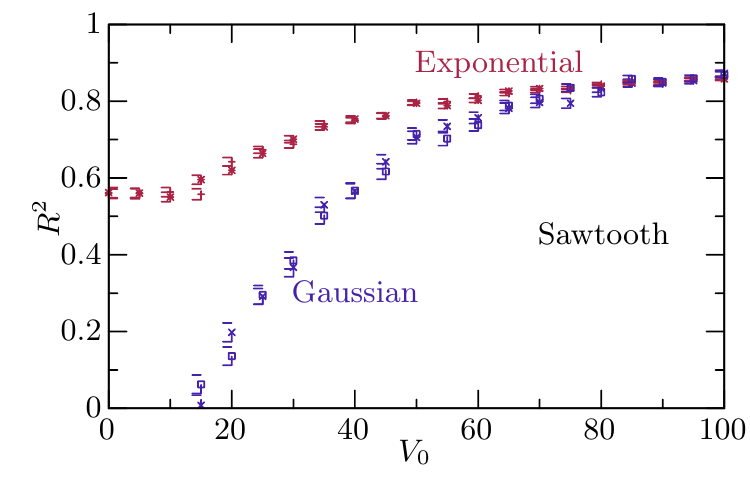}
	\vspace{0.2in}
	\includegraphics[width=0.9\linewidth]{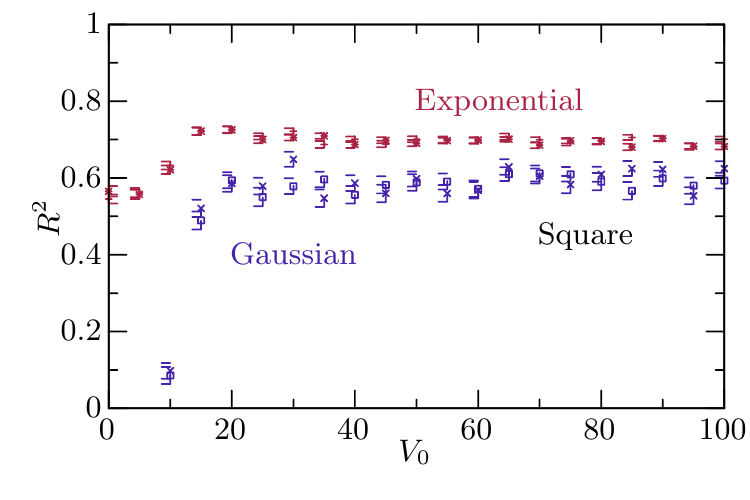}
	\caption{Plot of the coefficient of determination $R^2$ when fitting the absolute value of the lowest energy free state $\abs{\psi}$ to a Gaussian (\textcolor{bl}{blue}) or an exponential (\textcolor{re}{red}) as the depth of a confining potential $V_0$ is increased for a sawtooth (upper,~\eqnref{vsaw}) and square (lower,~\eqnref{vsq}) potential respectively. 2 sets of data are shown in each case to show the fit done in 2 dimensions. Wavefunction fitting done on a noninteracting system.}
	\label{formfitsq}
\end{figure}

In the non-interacting case, we solve for the lowest energy free state as in~\figref{formfit} to determine the shape of the localized wavefunctions. The results in~\figref{formfitsq} show that the sawtooth potential admits wavefunctions of indeterminate character while the square potential hosts wavefunctions that are definitively better described as exponential rather than Gaussian.

\begin{figure}[t]
	\centering
	\includegraphics[width=0.9\linewidth]{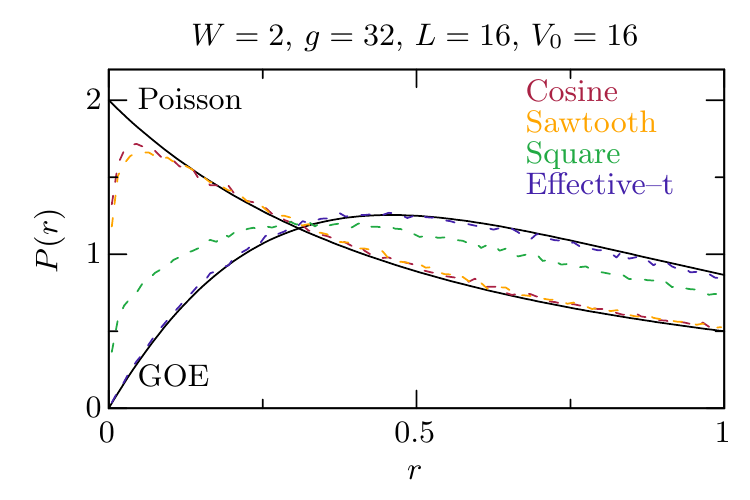}
	\caption{Plot of the distribution of level spacing ratios $r$ for an atomic gas under a cosine (\textcolor{re}{red},~\eqnref{mblh}), sawtooth (\textcolor{ye}{yellow},~\eqnref{vsaw}) and square (\textcolor{gr}{green},~\eqnref{vsq}) at fixed trap depth, and under no potential but with effective (reduced) hopping amplitude (\textcolor{bl}{blue},~\eqnref{effth}). Only the cosine and sawtooth potentials show appreciable agreement with the localized Poissonian prediction.}
	\label{lvspacsq}
\end{figure}

Turning on interactions, we see from~\figref{lvspacsq} that the system under the influence of the sawtooth potential is localized to a similar degree as that of the cosine potential, while the square potential does not appear to promote localization to the same degree, in agreement with our assertion on the importance of Gaussian localization.

{\it Hilbert space fragmentation: }Concomitant with application of a square potential in particular is the emergence of a new conserved quantity, that is the proportion of particles in the low (or high) potential regions, fragmenting the Hilbert space into weakly connected regions that become truly separate in the infinite $V_0$ limit. One must take care therefore when gathering level statistics as aggregating samples from disjoint sectors, that individually adhere to the GOE prediction, may result in a total sample more reminiscent of the Poisson distribution and thus a spurious indication of localization~\cite{univmoire}. We see in \figref{lvspacsq} that this has not happened, with the level statistics of a system under a square potential not being well described by the Poissonian prediction, in particular having no evidence of strong level repulsion. This is due to the new conserved quantity having a strong energy dependence and so by sampling in the middle of the spectrum, we exclusively sample the sector of Hilbert space that has half of the particles in the low potential region and half in the high potential region. Such emergent conserved quantities are not present in continuously varying potentials and so we conclude that the level statistics seen there are not due to spurious sampling of a fragmented Hilbert space.

{\it Effective hopping: }Another natural question to ask is whether the observed localization is a consequence of the potential energy term rescaling the kinetic hopping term, which was set to unity in \eqnref{mblh}. This idea is pertinent to the highly excited states we consider, where the particle(s) may be classically free and the (positive) difference between total and potential energy may be interpreted as kinetic energy. We start by considering the non-interacting Schr\"{o}dinger equation

\begin{align}
    E_n\psi_n(x)&=\left[-\nabla^2+V(x)\right]\psi_n(x)\nonumber\\
    &=-t(x)\nabla^2\psi_n(x),
\end{align}
and attempt to obtain a space-varying effective hopping $t(x)$ that encodes the effect of the applied potential $V(x)$ while maintaining the eigenstate $\psi_n(x)$ where $n$ is just some (set of) good quantum number(s) to label the state. The result is

\begin{align}\label{effectivet}
    t(x)=\frac{E_n}{E_n-V(x)},
\end{align}
and for $V(x)$ everywhere negative and focusing on classically free states, $E_n>0$, the effective hopping element is bounded as $0<t(x)<1$, indicating a reduced kinetic energy and thus, naively, an increased tendency for localization. We furthermore see that since $t(x)$ depends on $E_n$ and thus on the state labels $n$, there is in general no effective hopping model that preserves both the spectrum and eigenstates of the tight-binding model with confining potential. We can however select a particular $E_n$ to exactly recover one eigenstate and approximately recover others close in energy. To investigate this, we perform the same ED study on the effective-hopping Hamiltonian

\begin{align}
    H_t=&\sum_{i=1}^L\Bigg[ -t_i\left(c^\dagger_i c^{}_{i+1}+\mathrm{h.c.}\right)+w_in_i \nonumber\\&+\frac{g}{L}\sum_{i\neq j}\left(1-\frac{2\abs{i-j}}{L}\right)n_in_j\Bigg],\nonumber\\t_i=&\frac{E_{\mathrm{mid}}}{E_{\mathrm{mid}}-V_i},\label{effth}
\end{align}
where we now add interactions, $E_{\mathrm{mid}}$ is the energy of the exact eigenstate at the middle of the spectrum and the form of $t_i$ comes from~\eqnref{effectivet}. For the system parameters given, $E_{\mathrm{mid}}<0$ generically for any disorder configuration, and so with $V_i\geq0$, we have $0\leq t_i\leq1$. This would be expected to push the system towards localization, however as seen in the blue line of~\figref{lvspacsq}, $P(r)$ remains consistent with the GOE prediction even at high effective $V_0$ and so a simple rescaling of the hopping terms consistent with the applied potential $V_i$ is insufficient to trigger a transition to the MBL phase. 

\end{document}